\RequirePackage{fix-cm}
\documentclass[twocolumn,pdftex]{svjour3}          
\usepackage{graphicx}
\usepackage{subfigure}
\usepackage[ansinew]{inputenc}
\usepackage{amsmath}
\usepackage{numprint}
\usepackage[english]{babel}

\usepackage[authoryear]{natbib}

\journalname{Experiments in Fluids}
\begin{document}
\npthousandsep{}

\title{Particle tracking for polydisperse sedimenting droplets in phase separation}

\titlerunning{Particle tracking for polydisperse droplets} 

\author{T. Lapp \and M. Rohloff \and J. Vollmer \and B. Hof}

\authorrunning{T. Lapp et al.}

\institute{Max-Planck-Institute for Dynamics and Self-Organization \\
					 D-37073 Goettingen, Germany \\
           \email{tobias.lapp@ds.mpg.de}}

\date{Received: date / Accepted: date}

\maketitle

\begin{abstract}
When a binary fluid demixes under a slow temperature ramp,
nucleation, coarsening and sedimentation of droplets lead to an oscillatory evolution of the phase separating system. 
The advection of the sedimenting droplets is found to be chaotic. 
The flow is driven by density differences between the two phases.
Here, we show how image processing can be combined with particle tracking to resolve 
droplet size and velocity simultaneously. Droplets are used as tracer particles, and the sedimentation velocity is determined.
Taking these effects into account, droplets with radii in the range of 4 -- 40$\mu$m are detected and tracked. 
Based on this data we resolve the oscillations in the droplet size distribution which are coupled to the convective flow.
\end{abstract}

\section{Introduction}
\label{intro}

The characterization of particle distributions in fluids is important to control and optimize processes in food, pharmaceutical, 
oil and chemical industry, as \cite{Heffels1998} point out.
To determine the particle mass flux, particle sizes and velocities have to be measured simultaneously, which was achieved by \cite{Petrak2002}.

Here, we present a particle tracking algorithm, which uses droplets as marker particles to measure the flow field. 
They are created naturally in the phase separation process of demixing binary systems.
Droplet positions and radii are detected simultaneously. 
The radius can therefore be used as a criterion to identify droplets in subsequent images.
Assuming Stokes law the sedimentation velocity can be calculated from the droplet radius.
The droplet velocity is decomposed into sedimentation velocity and advection by the flow. 
By subtracting the sedimentation velocity from the droplet velocity, the advection of
\textit{all} droplets can be used to measure the flow field.
With the advective flow field and the sedimentation velocity of each droplet, its position in the next frame can be predicted
and compared to the image.

Previous studies have extensively used particle tracking velocimetry as a measurement technique to investigate turbulent flows carrying small particles
(\cite{Maas1993,Malik1993,Ouellette2006,Kreizer2010}). 
To that end monodisperse tracer particles are added to the fluid whose position is detected by image processing of high speed camera data. 
While this procedure is followed successfully for mono\-disperse particles in single-phase flows, 
it is not easily applicable for measuring flow patterns in demixing binary systems, which have been studied by \cite{Vollmer1997b,Vollmer2002} and \cite{Emmanuel2006}. 

Tracer particles cannot be added as they would act as nucleation centers, and therefore affect the
droplet number density. Further more, colloidal particles aggregate on the 
interfaces and change the growth and coalescence rate, as \cite{Thijssen2010} have shown. 
For a review of the stabilizing effect of colloidal particles in emulsions see \cite{Binks2002} and \cite{Aveyard2003}.

There also is a broad range of acoustic / electro-acoustic and optical techniques
for the measurement of droplet size distributions, an overview given by \cite{Maass2009}. 
Several laser based techniques measure chord lengths that have to be transfered into droplet size distributions (\cite{Hu2006}), 
e.g. focus beam reflectance measurement (FBRM, see e.g. \cite{Ruf2000}) or optical reflectance measurement (ORM, see e.g. \cite{Lovick2005}). 
However, \cite{Maass2011} and \cite{Simmons2000} have shown that these techniques give inaccurate results for liquid/liquid dispersions 
in comparison to image analysis of in situ microscope imaging. 

To investigate the evolution of demixing systems, it is therefore preferential to use particle tracking of droplets present in the demixing system, as described in the following.

The article is organized as follows:
We first characterize the chemical system. 
Then we introduce the experimental setup with a focus on the illumination technique, and explain the experimental procedure (sec. 2). 
In section 3, we present the particle tracking algorithm.
It is based on an image processing method to detect the droplets. 
In section 4, we measure the sedimentation velocity of droplets and give an indication on the uncertainties in the radius detection. 
We then use the velocity information to filter out falsely detected droplets.
We show how the droplet size distribution and the characteristics of the flow field evolve in time.
Finally, in section 5, we discuss the applicability and the limitations of the droplet tracking procedure for the investigation of demixing binary fluids.

\section{Experimental Method}

\subsection{Model System}

\begin{figure}[htb]
	\centering
	\includegraphics[width=0.8\linewidth]{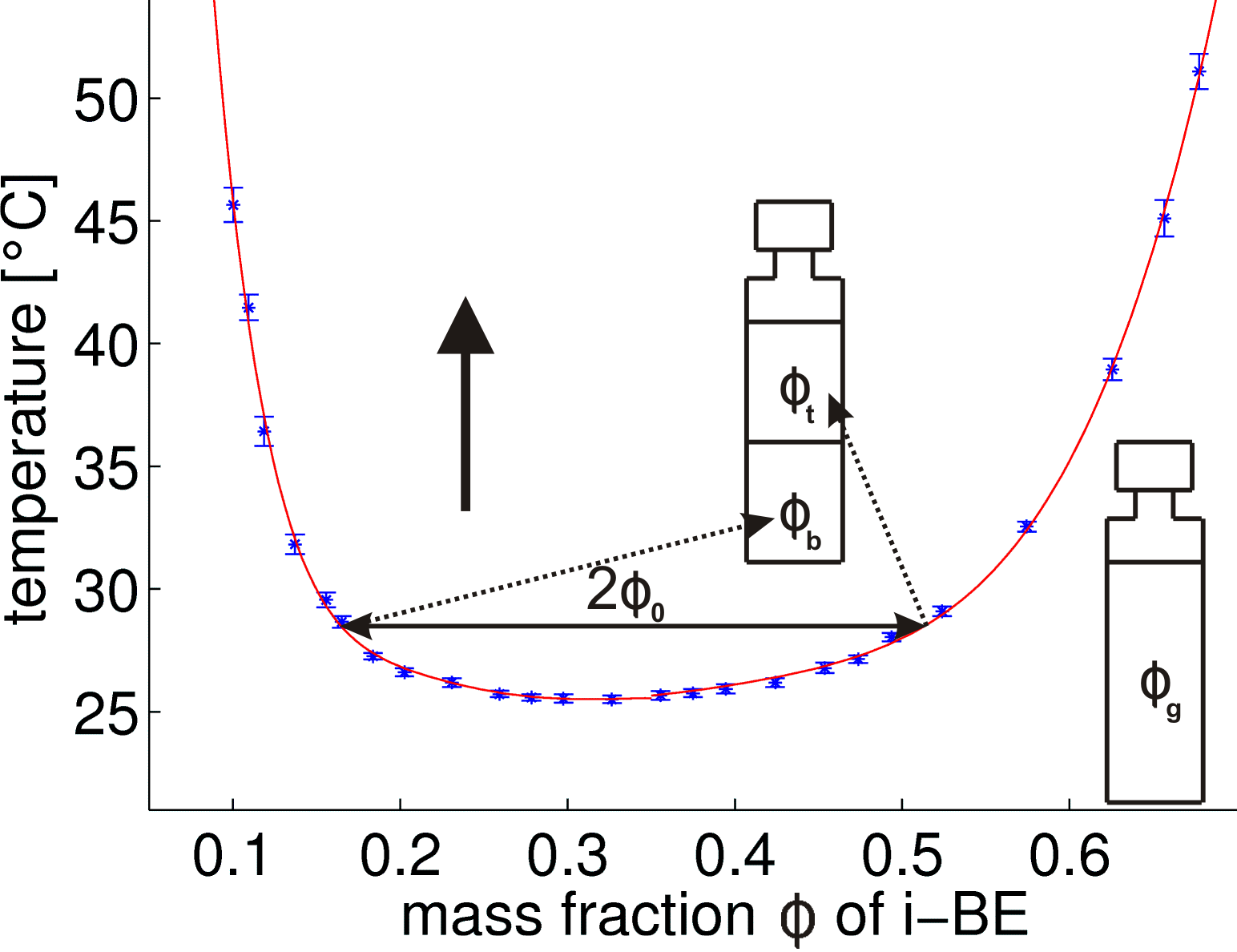}
	\caption{Phase diagram of water/isobutoxyethanol mixture between 25 and 50°C. 
		The samples are filled with the composition $\Phi_{\rm g}$.
		The mixture separates above a critical temperature $T_{\rm c} =$ 25.5°C.
		For slow temperature ramps the compositions $\Phi_{\rm b}, \Phi_{\rm t}$ of the bottom and top phase follow the left and right branch of the binodal, respectively.
		$2\Phi_{\rm 0}=\Phi_{\rm t}-\Phi_{\rm b}$ denotes the width of the miscibility gap.
		The left and the right branch are fitted with sixth and fourth order polynomials, respectively.}
	\label{fig:phase_diagram}
\end{figure}

As a model system a mixture of water and isobutoxy\-ethanol (i-BE) is chosen. 
It has a lower critical point at 25.5°C and demixes under heating.
A phase diagram is shown in Fig. \ref{fig:phase_diagram}.
The sample can easily be prepared at room temperature and demixes at temperatures between 25.5°C and 80°C, as shown by \cite{Nakayama2001}. 
The i-BE (purity $\geq 97 \%$) is purchased from Wako Chemicals GmbH and used without purification. 
The data points in the phase diagram (Fig. \ref{fig:phase_diagram}) were determined by turbidity measurements.
The solid lines show fits with sixth order (left branch) and fourth order (right branch) polynomials respectively.
 
The density and the viscosity of the mixture depend on composition as well as on temperature. 
For the density, the data of \cite{Doi2000} is used, taking thermal expansion and molar excess volume explicitly into account. 
The determination of the viscosity is described in the appendix.

Droplet detection is improved by using Nile Red as a fluorescent dye. 
Nile Red solved in butoxyethanol absorbs in the wavelength range of 500nm to 590nm (green light) and emits between 580nm and 700nm (red). 
The dye solves preferentially in organic compounds and poor\-ly in water. \cite{Fowler1985} point out that the fluorescence of Nile Red depends strongly on the
polarity of the solvent. In particlar, its fluorescence is quenched in water . 
Therefore, the i-BE-rich phase appears bright and the water-rich phase stays relatively dark. 
Since each phase consists of a mixture of water and i-BE, the contrast between dark and bright parts in the images recorded during a measurement depends on the composition of the two phases. 
It was checked that the dye has no significant influence on the phase diagram.

\subsection{Experimental Setup}

\begin{figure}[htb]
	\centering
	\includegraphics[width=0.8\linewidth]{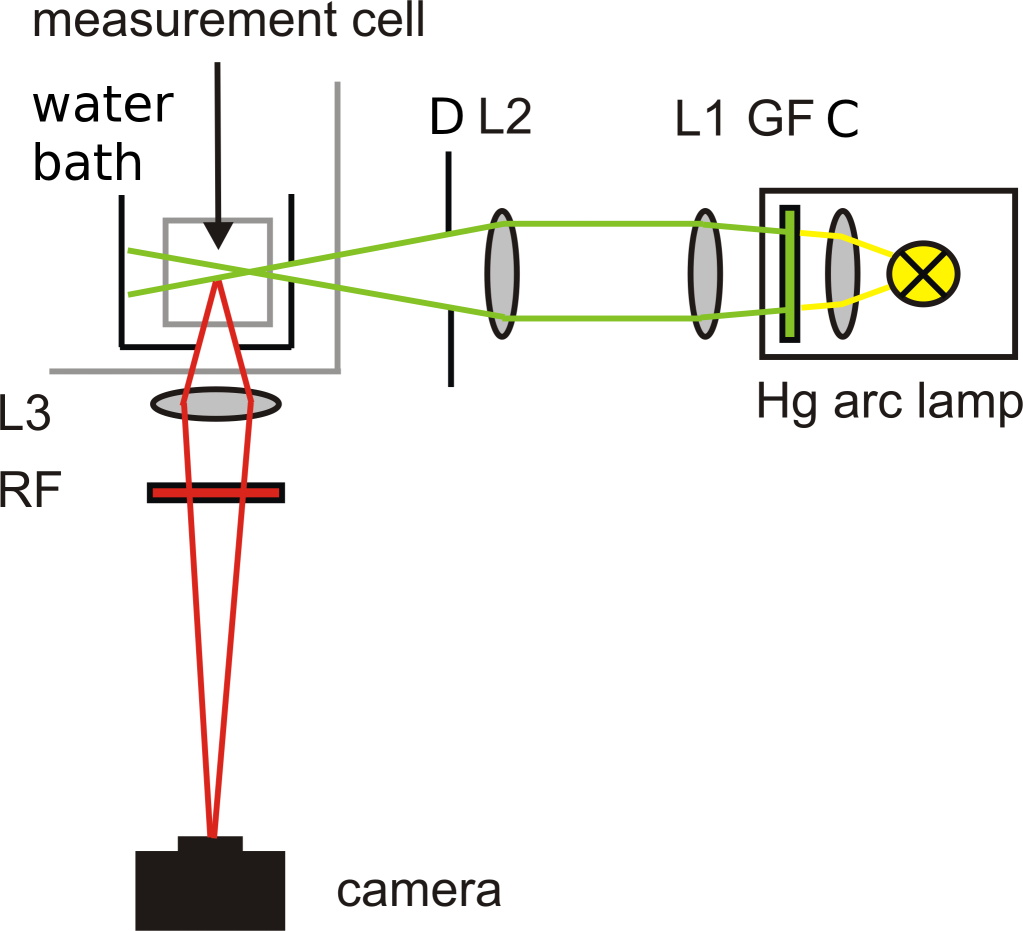}
	\caption{Sketch of the experimental setup, top view. 
		The light of a short arc Hg vapor lamp is collected by a collimator lens (C) and parallelized by a spherical lens (L1).
		A cylindrical lens (L2) forms a light sheet which is cut by a diaphragm (D) and a line aperture (300$\mu$m width) directly in front of the sample cell.
		The light sheet illuminates a plane of the binary mixture, a green filter (GF) selects the wavelength range suitable for excitation of the dye. 
		The emission of the fluorescently labeled phase is passes a red filter (RF) and is projected by an objective (L3) onto a 5 Mpixel CCD camera chip.
		The sample cell is mounted in a water bath with controlled temperature.}
	\label{fig:setup}
\end{figure}

The probe is contained in a fluorescence cell ($10\times10\times33$mm) 117.100F-QS made by Hellma GmbH. 
The measurement cell is mounted in a water bath with controlled temperature (see Fig. \ref{fig:setup}).
An immersion cooler Haake EK20 is cooling with constant power, and a temperature control module Haake DC30 is heating the water bath to a preset temperature.
Additionally the temperature of the water near the sample is measured with a PT100 temperature sensor. 
The temperature is controlled with an accuracy of 15 mK.

To get enough signal from the fluorescently labeled i-BE-rich phase, a bright light source is required. 
A laser light sheet is not suitable for illumination since it is unidirectional and the two phases are not index-matched. 
Each droplet acts as a little lens which focuses and diffracts the parallel laser light sheet. 
After passing a short distance of the sample, the droplets have transformed the uniform light sheet into a stripe pattern,
which illuminates the droplets very inhomogeneously. 
In contrast a mercury short arc lamp (LOT-Oriel, 100W) with a bright light emitting spot gives far better results. 
It provides a high light intensity, which is diffuse enough that almost no stripes appear (figure \ref{fig:raw_image}).

A sketch of the optical components is given in figure \ref{fig:setup}.
The light is collected by the collimator lens (C) and a bandpass filter (GF) selects the green emission lines (546 and 577/579nm) for the excitation of the fluorescent dye. 
The spherical lens (L1, $f=200\rm mm$) forms a parallel light beam. 
The cylindrical lens (L2, $f=80\rm mm$) focuses the light normal to the focal plane of the camera.
A spherical diaphragm (D) allows to adjust the amount of light manually.
With a narrow slit of 300$\mu$m width directly in front of the measurement cell a vertical light sheet is formed. 
The measurement cell is covered by black apertures to shield stray light. 
The fluorescent light from the illuminated plane of the sample is projected by a $f=35$mm objective (L3) to the chip of a BM-500CL monochrome progressive scan CCD camera. 
It takes 2058 x 2448 pixel images with a maximum frame rate of 15Hz, covering 1.3mm$\times$1.5mm of the sample (1.6 pixel/$\mu$m). 
With a longpass filter (RF, edge 594nm) the excitation light is filtered.

The camera, temperature control and magnetic stirrer are connected to a computer. 
Measurements are fully automated using a LABVIEW program. 

\subsection{Experimental Procedure}

The temperature protocol for our experimental procedure is adapted from \cite{Auernhammer2005}. 
After mixing the sample at 24°C for one hour, the temperature is set to $T=25.8$°C (0.3K above critical point) and the system equilibrates for four hours. 
Then a temperature ramp is run from 25.8°C to 50°C, which is designed to keep the volume flux between the phases constant. 

Following \cite{Auernhammer2005} and \cite{Cates2003} a driving parameter $\xi$ is defined
\begin{equation}
	\xi = \frac{1}{\Phi_{\rm 0}(T(t))}\frac{\partial \Phi_{\rm 0}(T(t))}{\partial t}
	\label{eqn:xi}
\end{equation}
with $2\Phi_{\rm 0}=\Phi_{\rm t}-\Phi_{\rm b}$, where $\Phi_{\rm t}$ and $\Phi_{\rm b}$ denote the compositions of the top and the bottom phase, 
respectively (compare Fig. \ref{fig:phase_diagram}). 
Hence, $2\Phi_{\rm 0}(T)$ is the width of the miscibility gap for a given temperature $T$.
The driving parameter $\xi$ is equal to the rate of droplet mass production in the two phases. 
Equation (\ref{eqn:xi}) is inverted to calculate a temperature ramp with fixed driving,
\begin{equation}
	T(t+\delta t) = T(t) + \frac{\xi \Phi_{\rm 0}(T(t))}{\partial_{T} \Phi_{\rm 0}(T(t))} \delta t.
	\label{eqn:tempramp}
\end{equation}
Equation (\ref{eqn:tempramp}) is integrated with \numprint{10000} uniformly distributed time steps in the range [$t_{\rm 0},t_{\textrm{end}}$] chosen such that $T(t_{\rm 0})=25.8$°C and $T(t_{\textrm{end}})=50.0$°C. 
To implement the temperature scan, the temperature of the thermostat is set with a rate of 1 Hz according to the calculated temperature ramp.

\section{Particle Detection and Tracking}

In this section we describe a particle detection and tracking algorithm for polydisperse sedimenting particles. 
Radii and positions are detected simultaneously. 
The first part describes an image processing algorithm which detects droplets in separate images. 
In the next step the flow field is estimated. 
Then the droplets are tracked through a series of images. 
Finally the velocity field is recalculated based on the droplet trajectories. 
This procedure allows us to sort out artifacts (caused e.g. by dirt or overlaps of droplet images) of the image processing, 
and to determine Lagrangian particle velocities. 
The image processing is done with the MATLAB Image Processing Toolbox. 
The data (typically \numprint{20000} images per measurement) is processed on a computer cluster. 
For simplicity we restrict our description to the detection of the fluorescently labeled droplets in the bottom phase. 
By inverting the images and the direction of gravity the same algorithm can be used for the dark droplets detected in the top phase.

\subsection{Processing of Single Images}

\subsubsection{Preprocessing}

\begin{figure}[htb]
	\centering
	\subfigure[\label{fig:raw_image}]
	{\includegraphics[width=\linewidth]{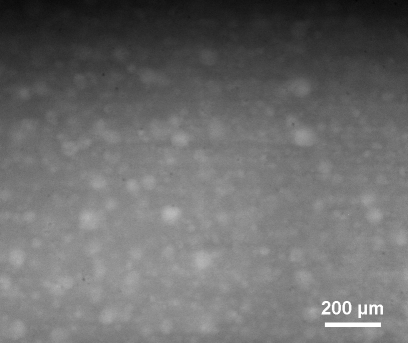}}
	\subfigure[\label{fig:preprocessed_image}]
	{\includegraphics[width=\linewidth]{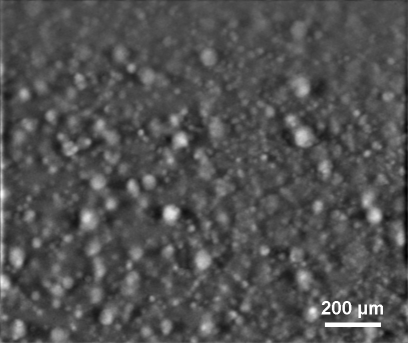}}
	\caption{(a) Raw image and (b) preprocessed image of i-BE-rich droplets in the water-rich bottom phase with $\xi = 1.05\cdot10^{-5}$s 
		 at $25.98\pm0.02$°C.
		 Horizontal stripes and the vertical intensity gradient are removed by Fourier filtering.}
	\label{fig:raw_prepro}
\end{figure}

In a first step, dark spots caused by dirt on the camera chip are removed with a flat field correction. 
Next the image is Fourier filtered and the contrast is optimized. 
To reduce the horizontal stripes (illumination from the right) produced by the light dispersion of the droplets 
a high pass filter (Gaussian filter with width 5 in horizontal and 200 in vertical direction) is applied.
For reducing the noise, a low pass filter (isotropic Gaussian filter, $\sigma = 100$) is applied. 
In figure \ref{fig:raw_prepro}, a raw image and a preprocessed image of fluorescent i-BE-rich droplets in the denser water-rich bottom phase are shown.
For high droplet densities, there is a significant number of overlaps of droplet images.
Therefore we need special algorithms to take care of this for detection.

\subsubsection{Droplet Detection}

\begin{figure}[htb]
	\centering
	\subfigure[\label{fig:thresholding0}]
	{\includegraphics[width=0.45\linewidth]{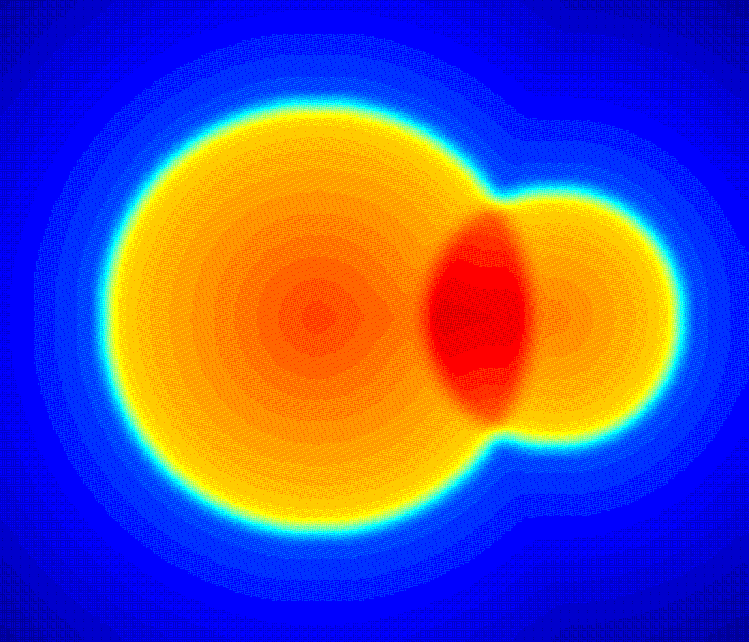}}
	\hfill
	\subfigure[\label{fig:thresholding1}]
	{\includegraphics[width=0.45\linewidth]{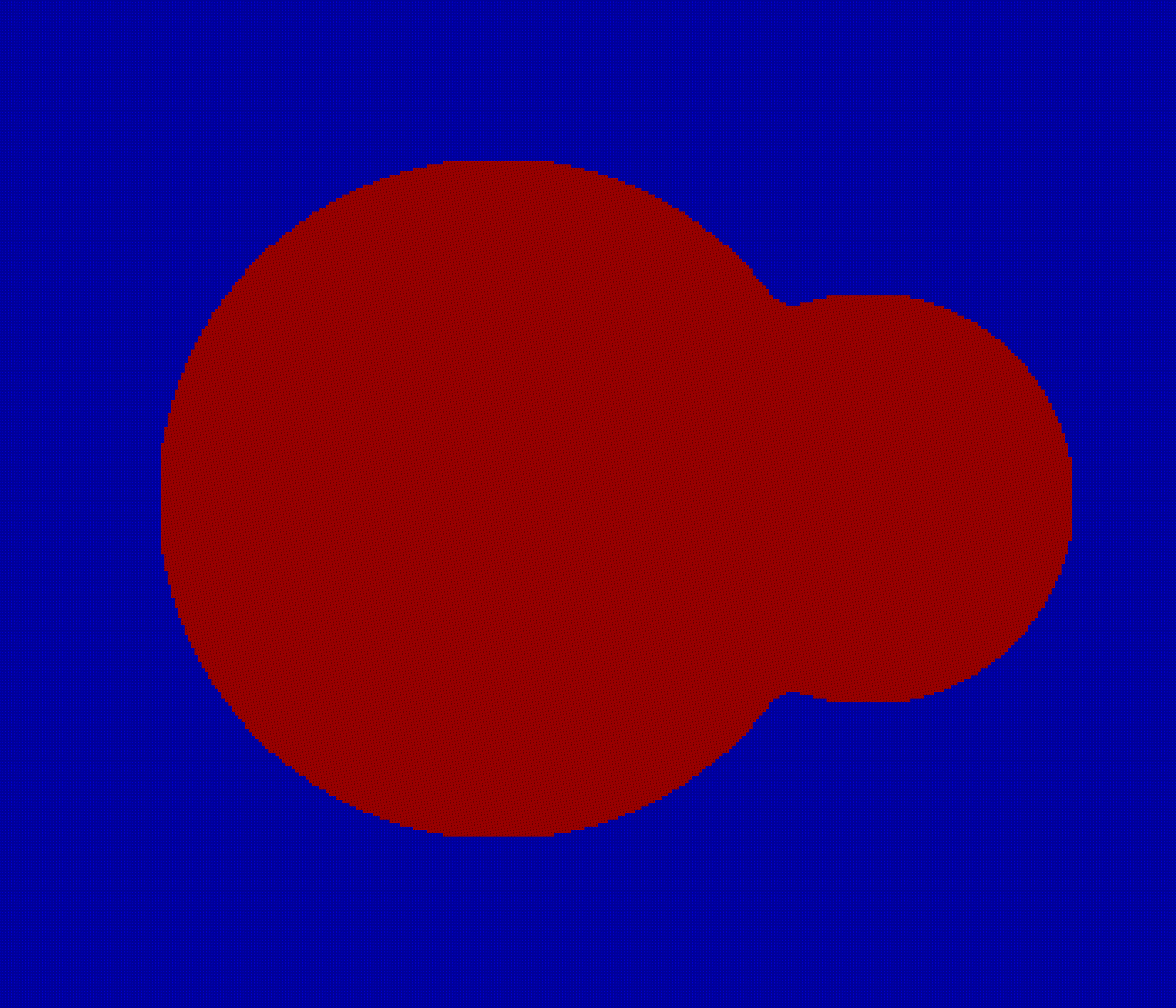}}
	\subfigure[\label{fig:thresholding2}]
	{\includegraphics[width=0.45\linewidth]{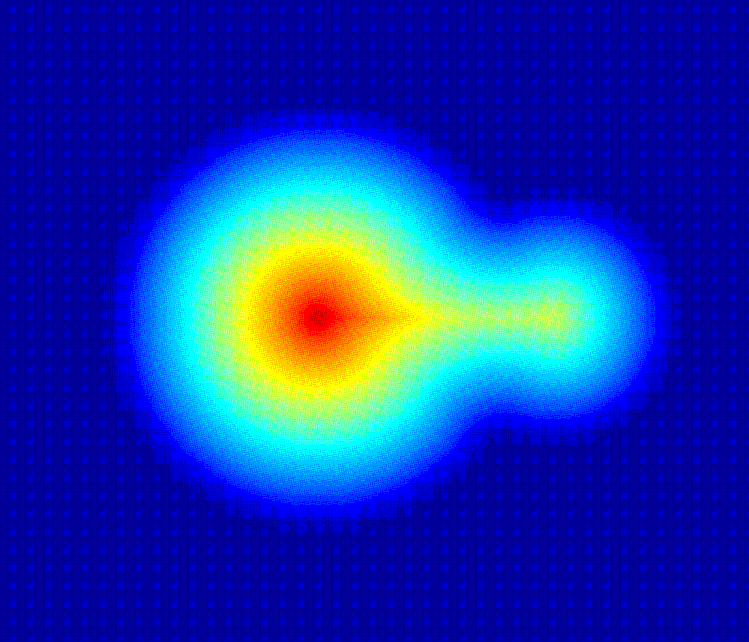}}
	\hfill
	\subfigure[\label{fig:thresholding3}]
	{\includegraphics[width=0.45\linewidth]{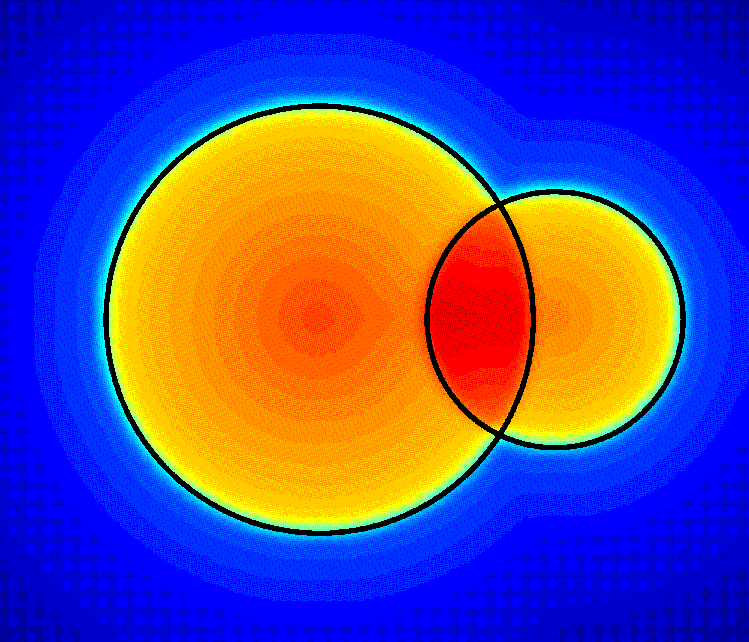}}
	\caption{Sketch of distance based image segmentation.\newline
					 (a) False color plot of two overlapping images of droplets.\newline
					 (b) Thresholded image.
					 (c) Distance from background.\newline
					 (d) Detected droplets (black) based on local maxima of distance.}
	\label{fig:thresholding}
\end{figure}

The droplets are detected in the preprocessed image using two strategies. 
The first relies on thresholding the image, the other uses marker-controlled watershed segmentation.
The first strategy is visualized in fi\-gure \ref{fig:thresholding} giving an example.
All pixels darker than a threshold level (Fig. \ref{fig:thresholding0}) are identified as background (Fig. \ref{fig:thresholding1}). 
Isolated background regions are deleted. 
Then, for each pixel the distance to the background is calculated (Fig. \ref{fig:thresholding2}). 
Local maxima in this distance map correspond to droplet centers, their distance to the background being their radius (Fig. \ref{fig:thresholding3}). 
This procedure is carried out twice with two thresholds, one being better for finding big droplets, the other for small droplets. 

\begin{figure}[htb]
	\centering
	\subfigure[\label{fig:watershed1}]
	{\includegraphics[width=0.45\linewidth]{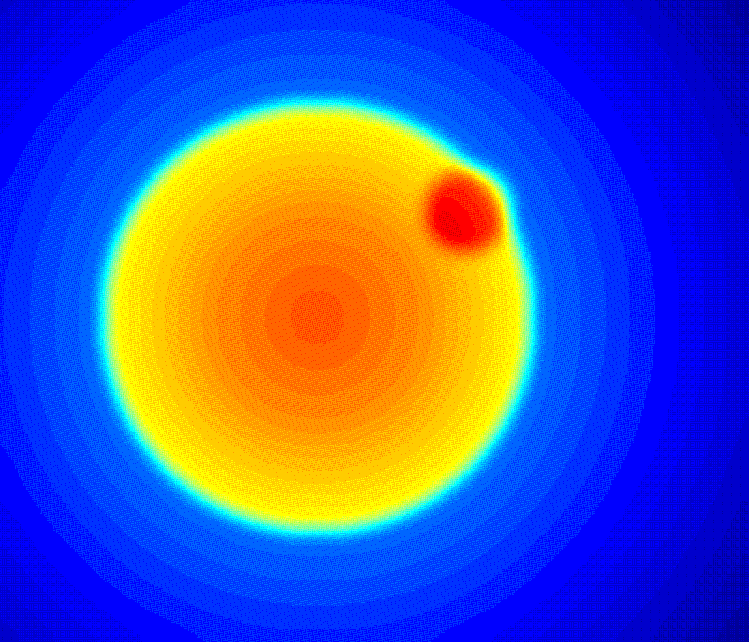}}
	\hfill
	\subfigure[\label{fig:watershed2}]
	{\includegraphics[width=0.45\linewidth]{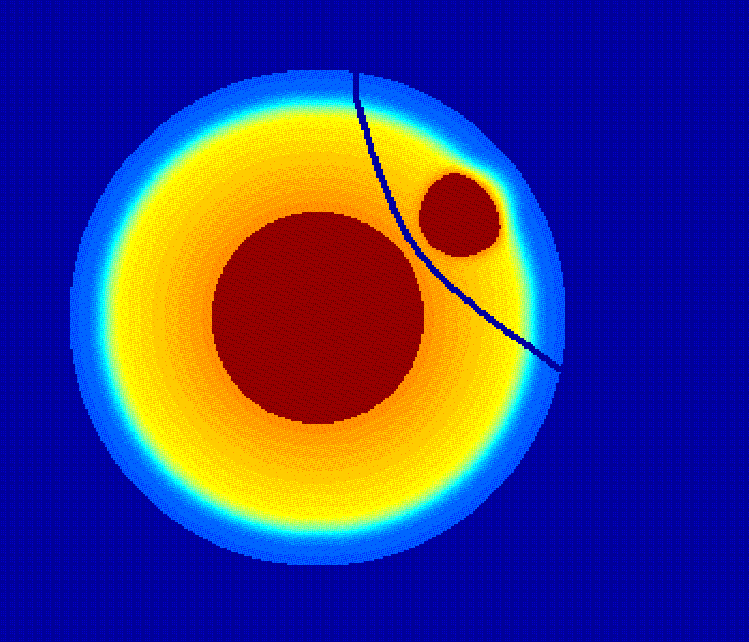}}
	\subfigure[\label{fig:watershed3}]
	{\includegraphics[width=0.45\linewidth]{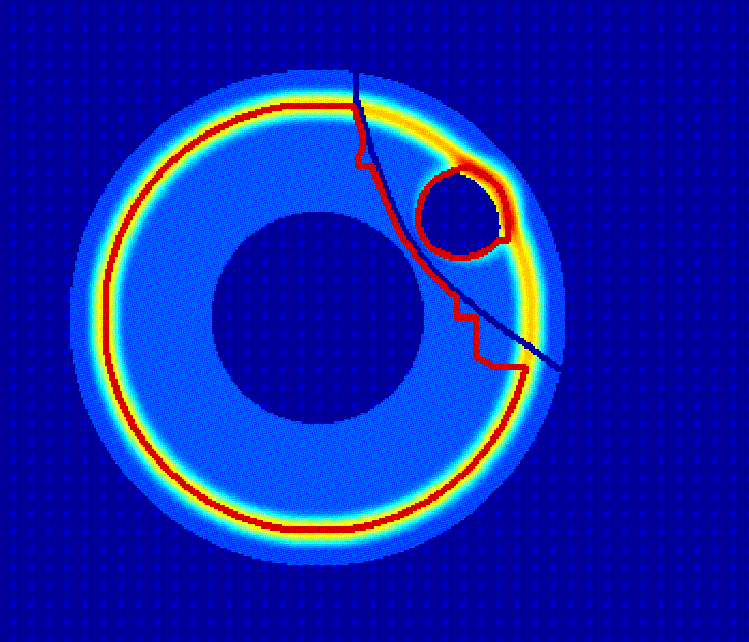}}
	\hfill
	\subfigure[\label{fig:watershed4}]
	{\includegraphics[width=0.45\linewidth]{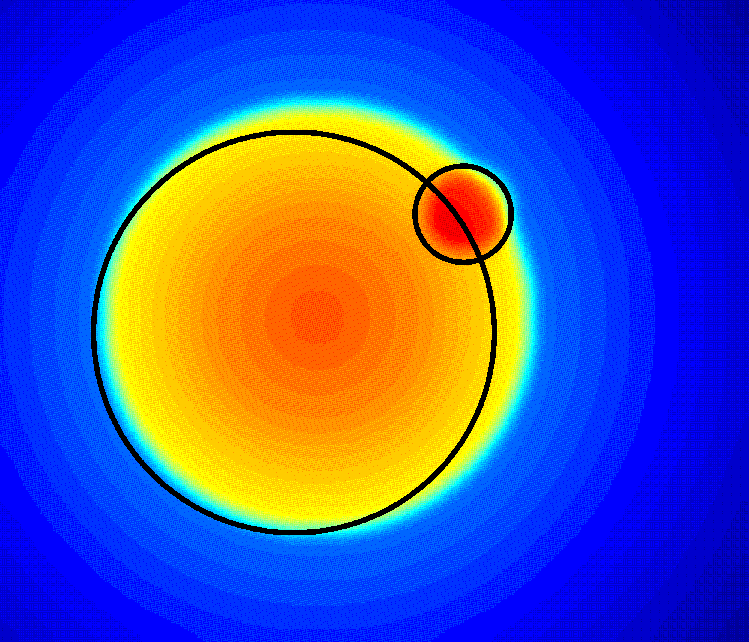}}
	\caption{Sketch of watershed image segmentation.
		(a) False color plot of two overlapping images of droplets.
		(b) Bright extended maxima mark droplets, lines in between and dark regions mark background.
		(c) Markers are imposed as minima (blue) on gradient image. Watershed lines follow droplet edges (red).
		(d) Detected droplets (black) based on regions enclosed by watershed lines.}
	\label{fig:watershed}
\end{figure}

The second strategy is based on the marker-control\-led watershed segmentation algorithm proposed by \cite{Gonzalez2004} pp. 422.
It is illustrated in figure \ref{fig:watershed}.
Extended maxima in the image (Fig. \ref{fig:watershed1}) are used as marker for the droplets (Fig. \ref{fig:watershed2}).
Lines between these maxima as well as extended minima are markers for the background. 
With a sobel filter the intensity gradient in the image is calculated (Fig. \ref{fig:watershed3}).
The markers are imposed as minima on the gradient image upon which a watershed transform is operated. 
The watershed lines of the intensity gradient correspond to droplet edges.
The markers are needed to avoid over-segmentation. 
Areas and centroids of the regions enclosed by the watershed lines are taken for the droplet radii and positions (Fig. \ref{fig:watershed4}).

Using these two approaches, even small droplets on top of bigger ones and overlapping images of droplets can be resolved. 
On the other hand, some droplets are found multiple times and some imaging artifacts are taken for droplets.
If two droplets with radii $r_{\rm 1},r_{\rm 2}$ are detected close to each other (centroid distance less than one radius) 
with almost the same radius (ratio $0.7<r_{\rm 1}/r_{\rm 2}<1.4$), 
they are considered as two detections of the same physical droplet.

\subsubsection{Matching to Image}

\begin{figure}[htb]
	\centering
	\subfigure[\label{fig:matchsketch_intensity}]
	{\includegraphics[width=0.45\linewidth]{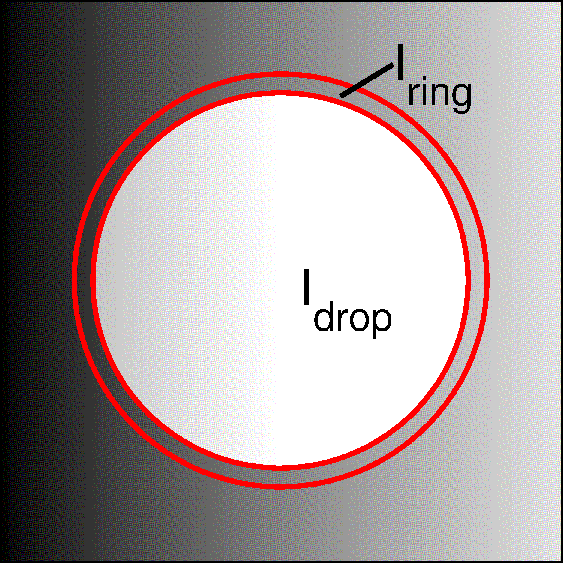}}
	\hfill
	\subfigure[\label{fig:matchsketch_centroid}]
	{\includegraphics[width=0.45\linewidth]{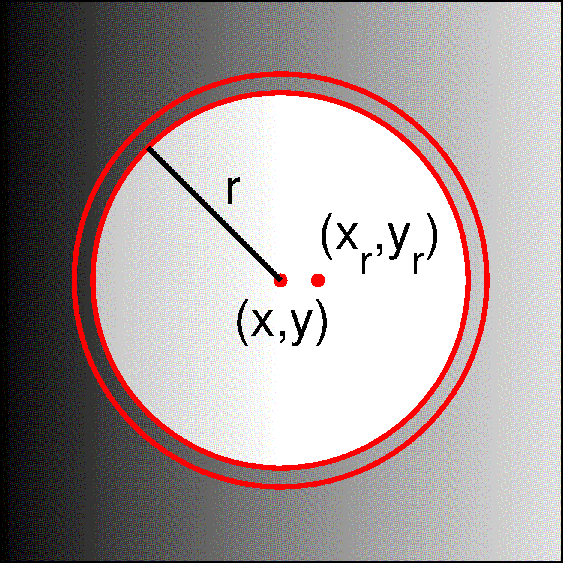}}
	\caption{Definition sketch for the match function.
					 (a) $I_{\textrm{drop}}$ denotes the mean intensity of the droplet, $I_{\textrm{ring}}$ the intensity of a ring around it.
					 (b) $(x,y)$ is the centroid of the droplet with radius $r$, $(x_{\rm r},y_{\rm r})$ is the intensity weighted centroid of the ring.}
	\label{fig:matchsketch}
\end{figure}

Not fully compensated dirt on the camera chip, overlaps of droplet images or unsharp droplet images can cause false detections.
Hence, a second step is implemented to distinguish droplets from artifacts and 
the estimates of the droplet radius $r$ and position $(x,y)$ [pixels] are compared to the preprocessed image. 
As characteristics of droplets in the image, we consider them being brighter than their environment and having a circular intensity gradient at the edge (figure \ref{fig:matchsketch}).
A match function $m$ is defined as
\begin{equation}
	\begin{aligned}
		m(x,y,r) = A(r) \left[I_{\textrm{drop}}(x,y,r)-I_{\textrm{ring}}(x,y,r)\right] \\
							 - B \frac{(x-x_{\rm r})^2+(y-y_{\rm r})^2}{r^2}
	\end{aligned}
	\label{eqn:matchfuntion}
\end{equation}
where the radius $r$ and the coordinates $x$, $x_r$, $y$ and $y_r$ are given in pixels. 
The difference between the mean intensity of the droplet $I_{\textrm{drop}}(x,y,r)$ and the mean intensity $I_{\textrm{ring}}(x,y,r)$ of a ring of 3 pixels width around it 
measures the intensity gradient at the droplet edge (figure \ref{fig:matchsketch_intensity}). 
An empirical radius dependent prefactor $A(r) = 1/\left(30 + r^{1.3}\right)$ ensures comparable matches for small and big droplets. 
Often small droplets are detected erroneously at the edge of big droplets. 
In this case, a little part of its halo is covering the dark background and the remaining part is covering the big droplet. 
The intensity weighted centroid of the halo (denoted as $x_{\rm r}$ and $y_{\rm r}$) is calculated and compared to the droplet position (figure \ref{fig:matchsketch_centroid}). 
The deviation accounts for this asymmetry and reduces the value of the match. It is weighted with the empirical prefactor $B=3$.  
For every droplet found in the droplet detection step, the match with the image is calculated. 
By varying radius and position of the droplet a local maximum of the match is found. 
All droplets with a match bigger than 0.08 are used for further analysis.

A final step to increase the reliability of the measured size distribution takes advantage of the time information of image sequences. 
Tracking the droplets allows to sort out unphysical trajectories of artifacts and to find undetected droplets in the image using its past or future trajectory.
To implement this information in section \ref{sec:filtering} we must evaluate the flow field and the droplet trajectories in the flow.

\subsection{Calculation of Flow Field}

\subsubsection{Identification of Droplets in Consecutive Images}

\begin{figure}[htb]
	\centering
	\includegraphics[width=\linewidth]{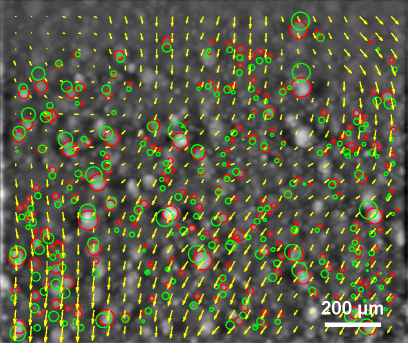}
	\caption{Fluorescently labeled droplets in the bottom phase (Fig. \ref{fig:preprocessed_image}). The arrows correspond to the flow field. 
					 Red circles mark the droplets found in the image and green circles their position and radius in the next image.}
	\label{fig:displfield}
\end{figure}

Typically droplets with radii $4\mu$m $<r<40\mu \rm m$ are detected. 
While small droplets ($r<10\mu$m) closely follow the flow, large droplets ($r>20\mu$m) are mainly driven by their buoyancy and do not qualify as tracers.
They sediment towards the interface due to the density difference of the two phases. 
In a Lagrangian frame, co-moving with the surrounding fluid, the sedimentation velocity $u_{\textrm{sed}}$ amounts to the Stokes velocity for a sphere of radius $r$, 
which is \cite[p. 234]{Batchelor}
\begin{equation}
	u_{\textrm{sed}} = \frac{2}{9} \frac{\Delta \rho  \: g r^2}{\eta}.
\label{eqn:vstokes}
\end{equation}
Here $\Delta \rho$ denotes the mass density difference between sphere and fluid, $g$ the gravitational acceleration, and $\eta$ the dynamic viscosity of the fluid.
Hence, the droplet velocity $u_{\textrm{drop}}$ can be split into a sedimentation and an advection term
\begin{equation}
	u_{\textrm{drop}}(x,y,r,t) = u_{\textrm{sed}}(r) + u_{\textrm{flow}}(x,y,t).
	\label{eqn:dropletvelocity}
\end{equation}
This ansatz will be checked in section \ref{sec:sedimentationvelocity}.
To track droplets the corresponding images of droplets in consecutive frames must be identified.
To this end, the velocity field of the previous time step $u_{\textrm{flow}}(x,y,t-\delta t)$ is taken as an initial guess for $u_{\textrm{flow}}(x,y,t)$.
For each droplet found in one image the position in the next image is predicted using equation (\ref{eqn:dropletvelocity}). 
The prediction is compared to the droplets found in the image, and the droplet pair with the closest distance between prediction and actual position 
and with similar radii is identified as one physical droplet. 
In figure \ref{fig:displfield} the droplets in the image are marked with red circles and the corresponding droplets in the next image with green circles.
The procedure used to determine the flow field, which is indicated by the arrows, is lined out in section \ref{sec:flowfield}.
Subsequently in section \ref{sec:particletracking} we describe details of the scheme adopted for finding particle trajectories.

\subsubsection{Determine the Flow Field}
\label{sec:flowfield}

In order to calculate the Eulerian flow field $u_{\textrm{flow}}(x,y,t)$ the image is sampled by an equidistant mesh with $25\times20$ grid points.
The grid distance is about $60\mu\rm m$, which is in the order of the droplet distance.
The correlation length of the flow field is about $200\mu \rm m$, so the grid can resolve the flow structures. 
A flow correlation time of 30s allows for small frame rates (order of 1Hz).
By subtracting the sedimentation velocity $u_{\textrm{sed}}$ from the droplet velocity $u_{\textrm{drop}}$ the flow contribution $u_{\textrm{flow}}$ can be calculated:
The advection of the $N$ droplets in each mesh cell $(i)$ is averaged, incorporating with half weight the $M$ droplets of neighboring cells $(ij)$:
\begin{equation}
    \begin{aligned}
	u^{(i)}_{\textrm{flow}} = \frac{1}{N+M/2} [ \sum_{k=1}^{N}(^{k}u^{(i)}_{\textrm{drop}}- ^{k}u^{(i)}_{\textrm{sed}}) \\
			     + \frac{1}{2} \sum_{k=1}^{M} \sum_{j=1}^{4}(^{k}u^{(ij)}_{\textrm{drop}}- ^{k}u^{(ij)}_{\textrm{sed}}) ]
    \end{aligned}
\end{equation}

\subsubsection{Smoothing of Flow Field}

The advection field $u^{(i)}_{\textrm{flow}}$ is smoothed with a weighted average of next-nearest neighboring cells to eliminate unphysical discontinuities.
\begin{equation}
	\bar{u}^{(i)}_{\textrm{flow}} = \frac{1}{4} u^{(i)}_{\textrm{flow}}
				   + \underbrace{\frac{1}{8} \sum_{j=1}^{4} u^{(ij)}_{\textrm{flow}}}_{\textnormal{next neighbours}} 
				   + \underbrace{\frac{1}{16} \sum_{j'=1}^{4} u^{(ij')}_{\textrm{flow}}}_{\textnormal{next-nearest neighbours}}
\end{equation}
In figure \ref{fig:displfield} a smoothed advection field is shown together with the matched droplets. 
The Eulerian flow field is used to predict the droplet positions in the next image. 
Note that the spatial distribution of detected droplets is not homogeneous and some grid points can only be calculated by interpolation. 
For the subsequent analysis of the flow field only the grid points which are directly computed from droplet displacements are taken into account.

\subsection{Particle Tracking}
\label{sec:particletracking}

\subsubsection{Forward Tracking}

\begin{figure}[htb]
	\centering
	\subfigure[\label{fig:forward1}]
	{\includegraphics[width=0.45\linewidth]{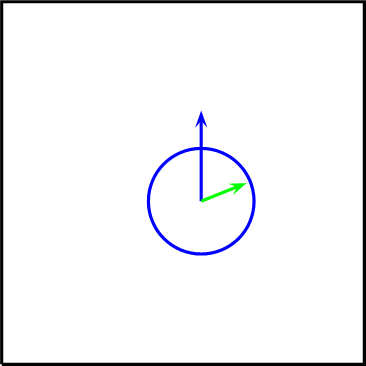}}
	\hfill
	\subfigure[\label{fig:forward2}]
	{\includegraphics[width=0.45\linewidth]{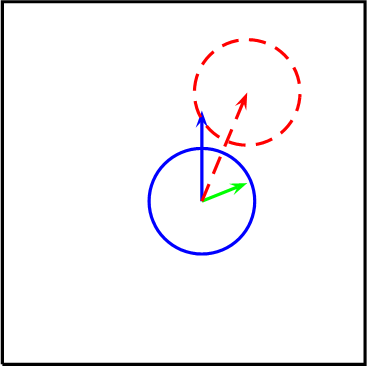}}
	\subfigure[\label{fig:forward3}]
	{\includegraphics[width=0.45\linewidth]{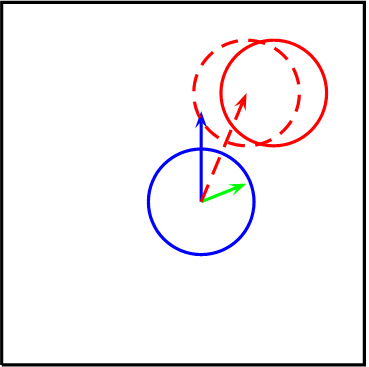}}
	\hfill
	\subfigure[\label{fig:forward4}]
	{\includegraphics[width=0.45\linewidth]{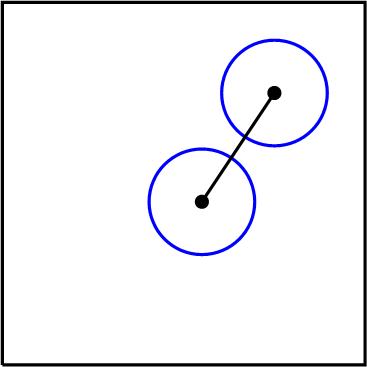}}
	\caption{Forward tracking: (a) Sedimentation velocity (blue, upward) and advection (green, 2 o'clock), 
				  (b) predict the next position (red dashed), 
				  (c) match the prediction to the next image (red solid) and
				  (d) build trajectory (black).}
	\label{fig:forward}
\end{figure}

Now that the advection field and sedimentation velocities for each droplet found in the image at time $t$ have been determined,
the position at time $t+\delta t$ can be predicted. 
This prediction is matched to the image of $t+\delta t$ and a local maximum of the match is searched for. 
This procedure is illustrated in figure \ref{fig:forward}. 
Having determined the displacement from time $t$ to $t+\delta t$, the advection field can be recalculated. 
In the next step, the positions of droplets tracked from image $t$ to $t+\delta t$ plus all other droplets 
also detected in image $t+\delta t$ are predicted for the time step $t+2\delta t$. 
They are compared to the droplets in image $t+ 2\delta t$ and the procedure is repeated. 
The identified droplet pairs are sorted into trajectories, which contain the position and radius values for the different frames.
Based on the trajectories, the flow field $u_{\textrm{flow}}(x,y,t)$ is recalculated according to the procedure described in section \ref{sec:flowfield}.

\subsubsection{Backward Tracking}

\begin{figure}[htb]
	\centering
	\subfigure[\label{fig:backward1}]
	{\includegraphics[width=0.45\linewidth]{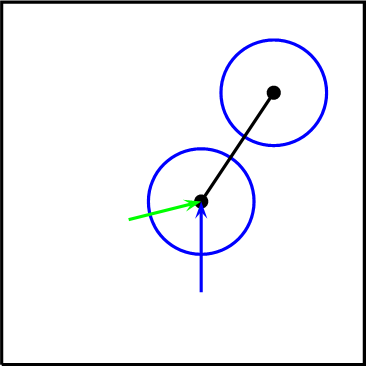}}
	\hfill
	\subfigure[\label{fig:backward2}]
	{\includegraphics[width=0.45\linewidth]{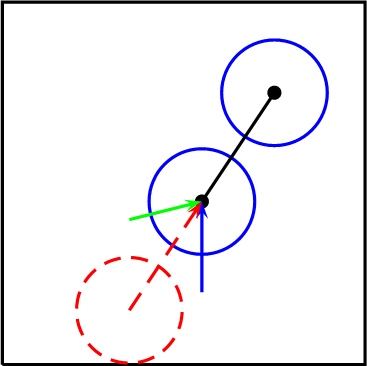}}
	\subfigure[\label{fig:backward3}]
	{\includegraphics[width=0.45\linewidth]{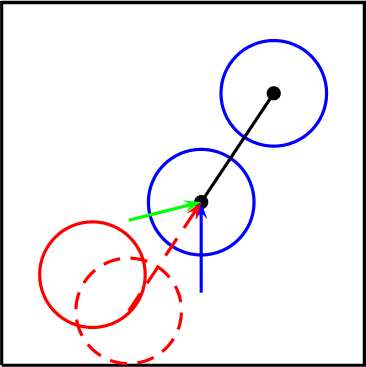}}
	\hfill
	\subfigure[\label{fig:backward4}]
	{\includegraphics[width=0.45\linewidth]{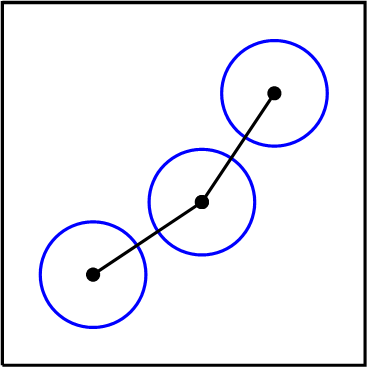}}
	\caption{Backward tracking: (a) Sedimentation velocity (blue,upward) and advection (green, 8 o'clock),
				    (b) predict the previous position (red dashed), 
				    (c) match the prediction to the previous image (red solid) and
				    (d) enlarge the trajectory (black).}
	\label{fig:backward}
\end{figure}

The forward tracking result can be improved by subsequently tracking the droplets backwards in time. 
The positions of droplets at the beginning of the trajectories are predicted for the time step before and matched to the image (compare figure \ref{fig:backward}). 
The droplet is compared to the entries in the other trajectories to merge interrupted trajectories, and to suppress branching of trajectories. 
As yet, our tracking algorithm does not account for the collision of droplets. 

\subsubsection{Filtering of Trajectories}
\label{sec:filtering}

Trajectories shorter than three time steps are removed.
Finally, all droplets which have moved by less than half the average advection velocity during the tracked time interval are deleted.
Most likely, those spots are imaging artifacts.
Eventually, the flow field is recalculated once more based on the enlarged and filtered trajectories in order to arrive at a consistent data set.

\section{Results}

\subsection{Sedimentation Velocities}
\label{sec:sedimentationvelocity}

\begin{figure}[htb]
	\centering
	\includegraphics[width=0.8\linewidth]{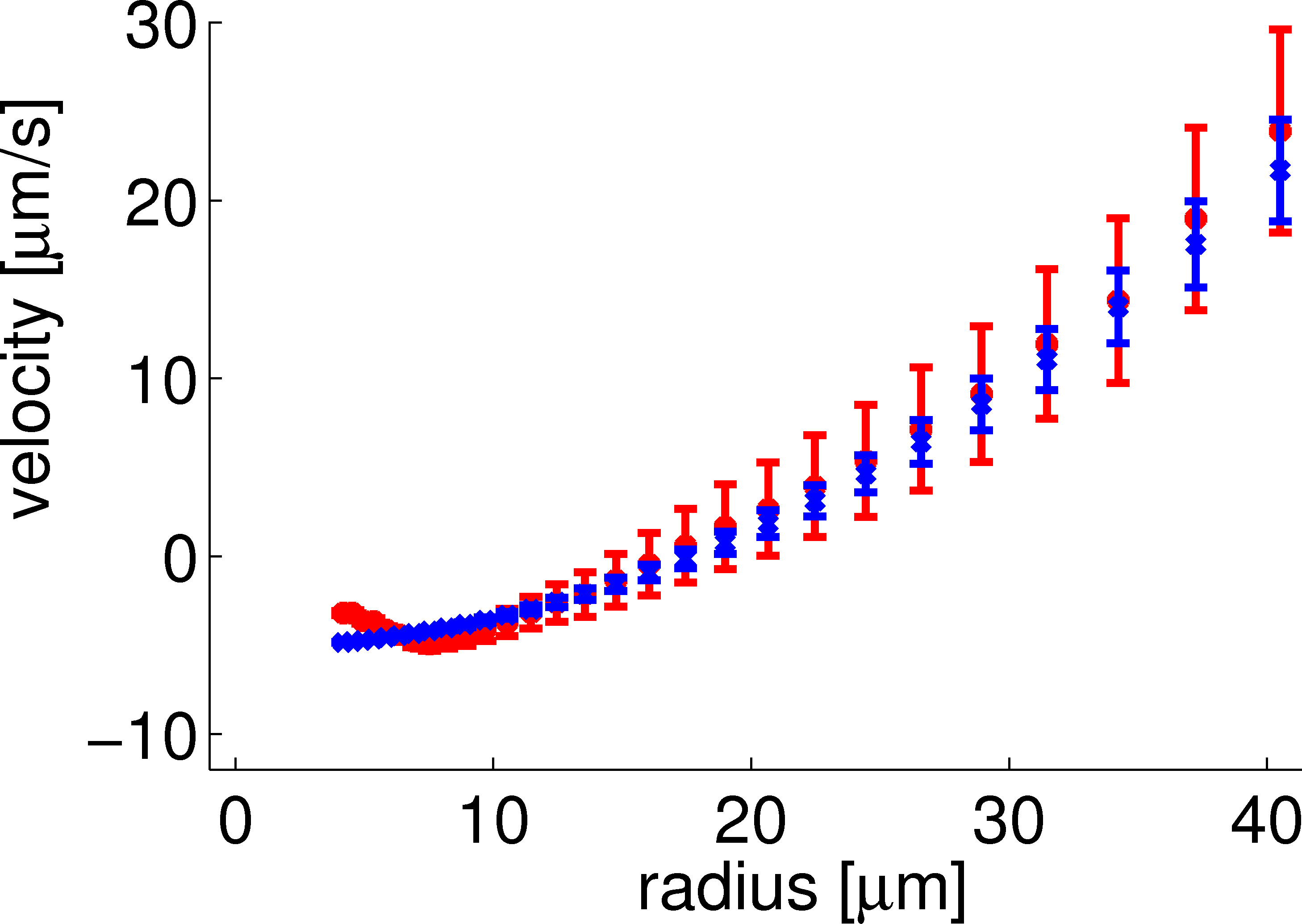}
	\caption{Comparison of the measured vertical velocity of the droplets (red) and the Stokes velocity modified by the mean flow (blue), in the bottom phase at 26°C.}
	\label{fig:velo_sed}
\end{figure}

We first aim to confirm the validity of the ansatz (\ref{eqn:vstokes}) and (\ref{eqn:dropletvelocity}).
In figure \ref{fig:velo_sed} the vertical velocity calculated from the trajectories of the tracked droplets is shown as a function of the radius (red data points). 
The error bars display the uncertainty of the radius detection (see below) and the change of the material parameters in the time interval of averaging. 
It is compared to the sedimentation velocity (blue points) due to equation \ref{eqn:vstokes} modified by the mean vertical velocity of the flow, which is calculated from the flow field. 
The Stokes velocity of a sphere (\ref{eqn:vstokes}) appears to describe the average settling rate very well. 
We attribute the deviations for small droplet radii to the fact that both the droplet size distribution and the average flow velocity oscillate. 
Most of the small droplets can be found at the beginning of the oscillations, when the average downward flow velocity is relatively small (see below).

\subsection{Stokes Number}

Equation (\ref{eqn:dropletvelocity}) holds for droplets with a Stokes number $\textrm{St}\!\ll1$. 
The Stokes number is defined as the ratio of two time scales
\begin{equation}
	\textrm{St} = \frac{\tau_{\textrm{drop}}}{\tau_{\textrm{flow}}}.
\end{equation}
In a flow which changes on the time scale $\tau_{\textrm{flow}}$ the advected droplets respond to accelerations on the time scale
\begin{equation}
	\tau_{\textrm{drop}}\equiv \frac{2 \Delta \rho \: r^{2}}{9 \eta}.
\end{equation}
The correlation time of the flow is found to be $\tau_{\textrm{flow}}\sim 30\rm s$. 
As an upper limit for the droplet response time we find $\tau_{\textrm{drop}}=36\mu \rm s$ at $T=50$°C for droplets with $a=40\mu\rm m$.
Therefore, typical Stokes numbers in the experiment are smaller than $10^{-6}$.
Even for a flow with high turbulence intensity with a Kolmogorov time scale of the order of microseconds,
the Stokes number would not exceed $10^{-3}$.
Due to the small density differences of fluids, kinematic particle-turbulence interactions (for a review see \cite{Vaillancourt2000}) 
can hence be ruled out for droplets, even in a turbulent flow. 

\subsection{Trajectory Length}

\begin{figure}[htb]
	\centering
	\includegraphics[width=0.8\linewidth]{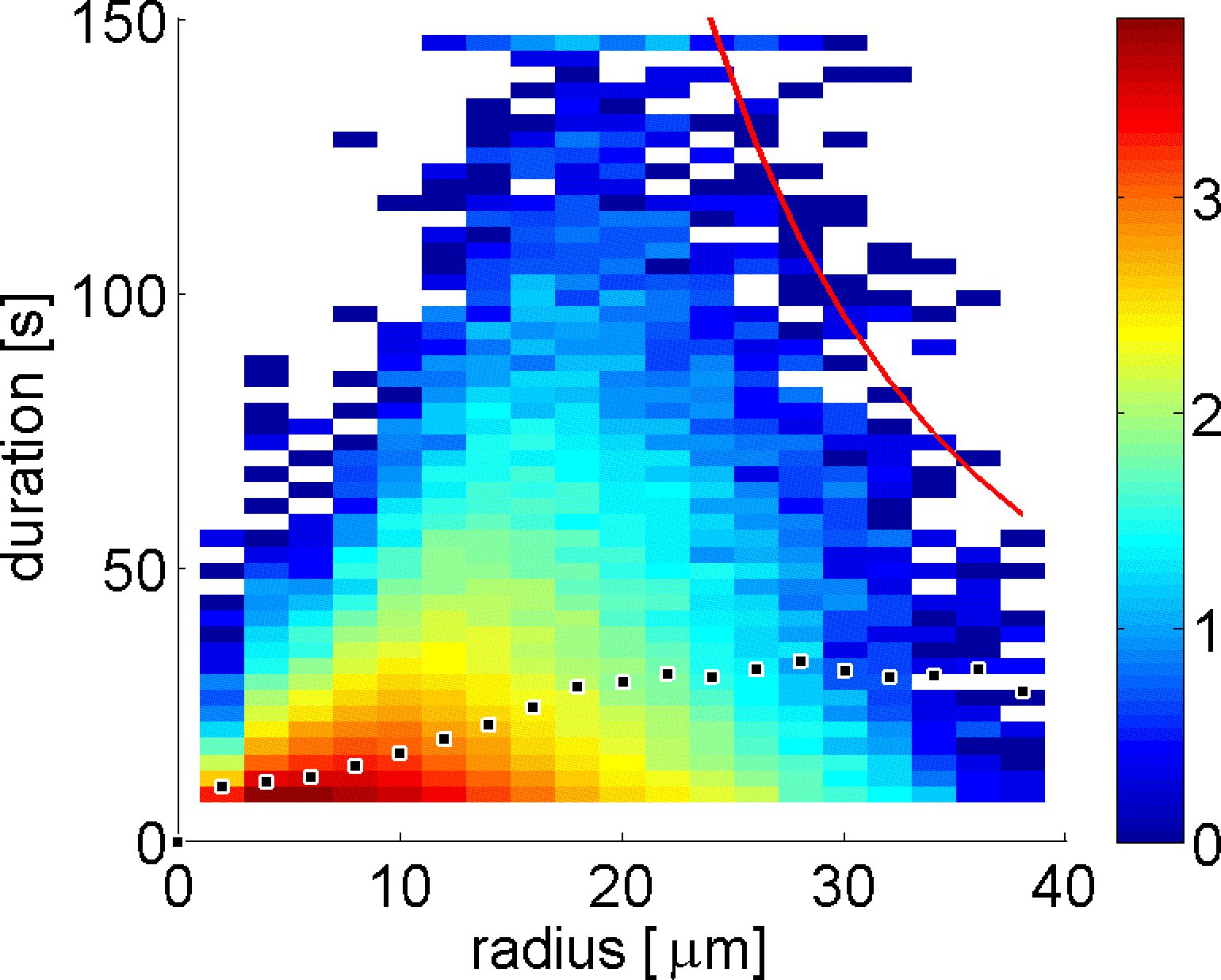}
	\caption{Radius dependence of the trajectory length: The number of trajectories with a given radius and length are color coded on a logarithmic scale. 
					 The black squares denote the average length for each radius bin and the solid red line is the sedimentation boundary.}
	\label{fig:traj_duration}
\end{figure}

For a quantitative analysis trajectories of 2000 images are sorted into radius bins according to their average radius. 
In each bin the number of trajectories of a given length is color coded on a decade logarithmic scale (figure \ref{fig:traj_duration}). 
The red line indicates the time $t = H/u_{sed}$ needed by a droplet settling with Stokes velocity to pass through the image of height $H$. 
It puts an upper bound on the trajectory length and leads to a decrease of the trajectory length for big radii.

For small radii, the duration of individual droplet detection is observed to increase roughly linearly with the radius (figure \ref{fig:traj_duration}). 
Hence we assume that the depth, in which a droplet can be detected, is proportional to its radius. 
Small droplets have to be exactly in the focal plane of the camera to be detectable, where as big droplets can still be detected when their center is slightly off the focal plane. 
Since the radius is detected by the maximum of the intensity gradient, it is rather robust to slight blurring,
and once the droplets are too far from the focal plane, they are no longer detected because the intensity gradient at their edge is below the threshold of the match function.

\subsection{Uncertainty in Radius Detection}

The fluctuations of the detected droplet radii in the trajectories are calculated to estimate the uncertainty of the attributed radius. 
For radii $r\leq19\mu$m a linear increase of the standard deviation is observed, yielding a constant relative uncertainty of about 20\%. 
Above 19$\mu$m, the standard deviation is more or less constant with $3.8 \mu$m.

\begin{figure}[htb]
	\centering
	\includegraphics[width=0.8\linewidth]{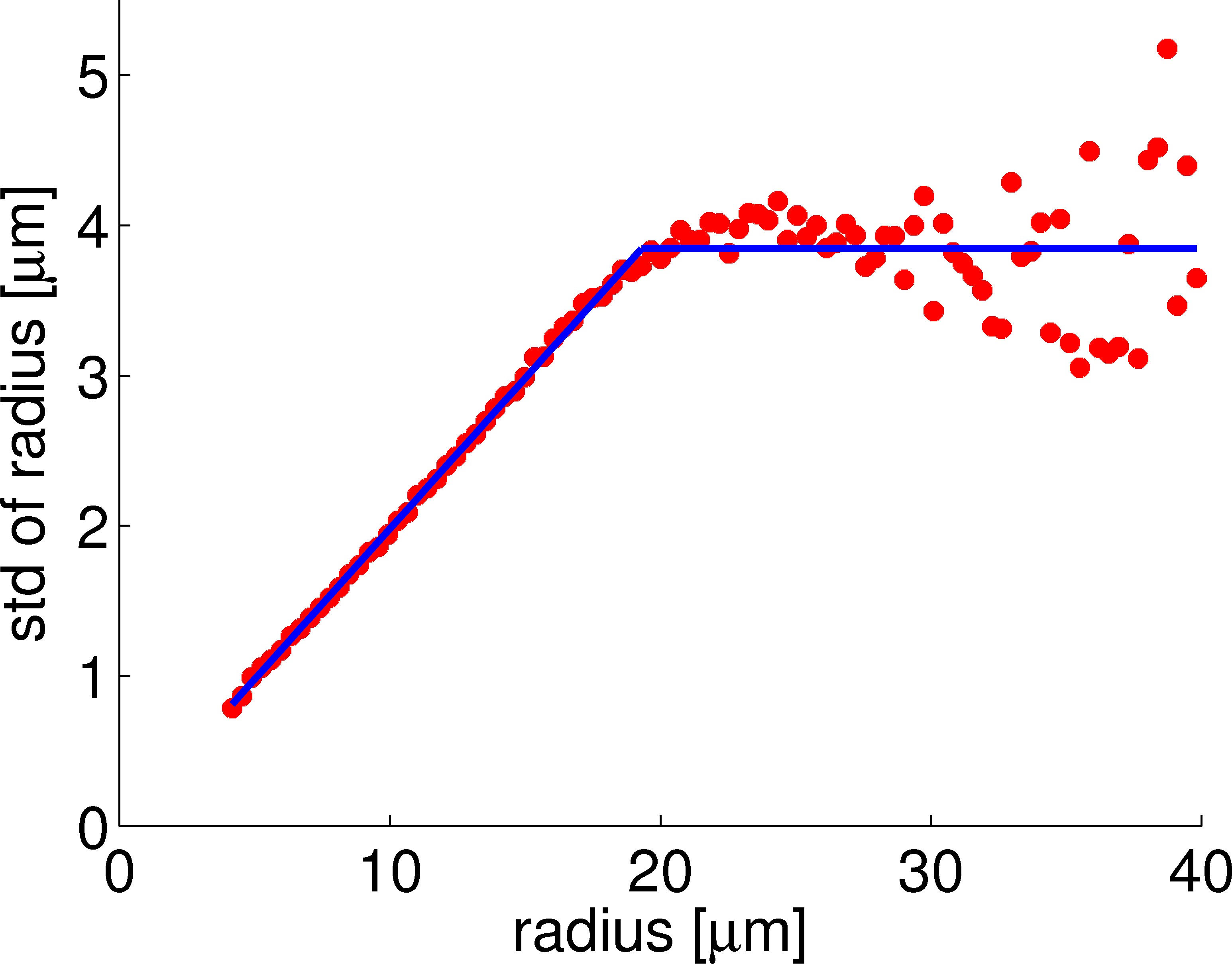}
	\caption{Radius-dependent standard deviation of the radius of tracked droplets..}
	\label{fig:radius_fluct}
\end{figure}

\subsection{Oscillating Size Distribution of Droplets}

\begin{figure}[htb]
	\centering
	\subfigure[\label{fig:trhistogram_ip}]
	{\includegraphics[width=0.8\linewidth]{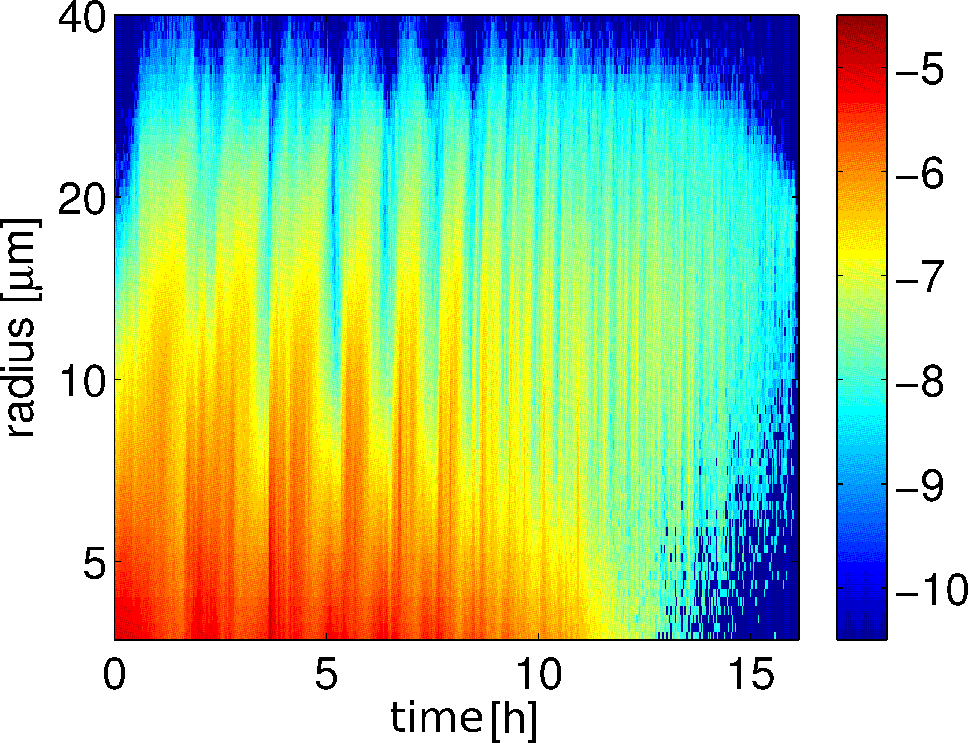}}
	\hfill
	\subfigure[\label{fig:trhistogram_tr}]
	{\includegraphics[width=0.8\linewidth]{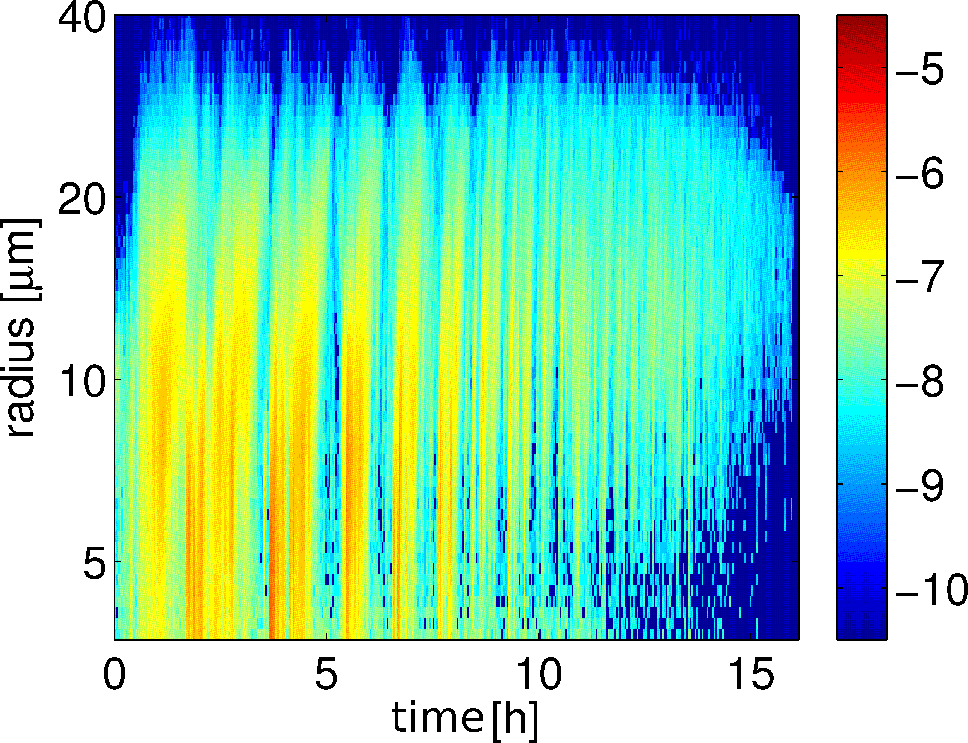}}
	\caption{Size distribution of droplets of the bottom phase with $\xi = 1.05\cdot 10^{-5} $s$^{-1}$: 
					(a) before and (b) after consistency check based on droplet tracking. 
					The droplet number density per radius given in $\mu$m$^{-4}$ is color coded on a decadic logarithmic scale.}
	\label{fig:sizedistribution}
\end{figure}

By slowly heating the system, more than twenty oscillations in the size distribution can be found (Fig. \ref{fig:sizedistribution}). 
A complex interplay of nucleation (\cite{Krishnamurthy1980,Sagui1999,Vollmer2008}), coarsening (\cite{Aarts2005,Vollmer2007}) and sedimentation (\cite{Hayase2008}) 
causes oscillations of the supersaturation and turbidity, modeled by \cite{Vollmer2007} and \cite{Benczik2010}.
They can also clearly be identified in false color plots of the droplet size distribution displayed in Fig. \ref{fig:sizedistribution}.
Comparing the two size distribution plots in Fig. \ref{fig:sizedistribution} reveals the effectiveness of droplet tracking. 
The noise in the processing of single images is reduced significantly and the evolution of the size distribution is unraveled for small droplets also. 
A detailed analysis of the oscillating size distribution and its implications for the understanding of precipitation in other systems 
(for example rain formation, for a review see \cite{Shaw2003}) will be given elsewhere.

\subsection{Oscillating Flow Properties}

\begin{figure}[htb]
	\centering
	\includegraphics[width=0.8\linewidth]{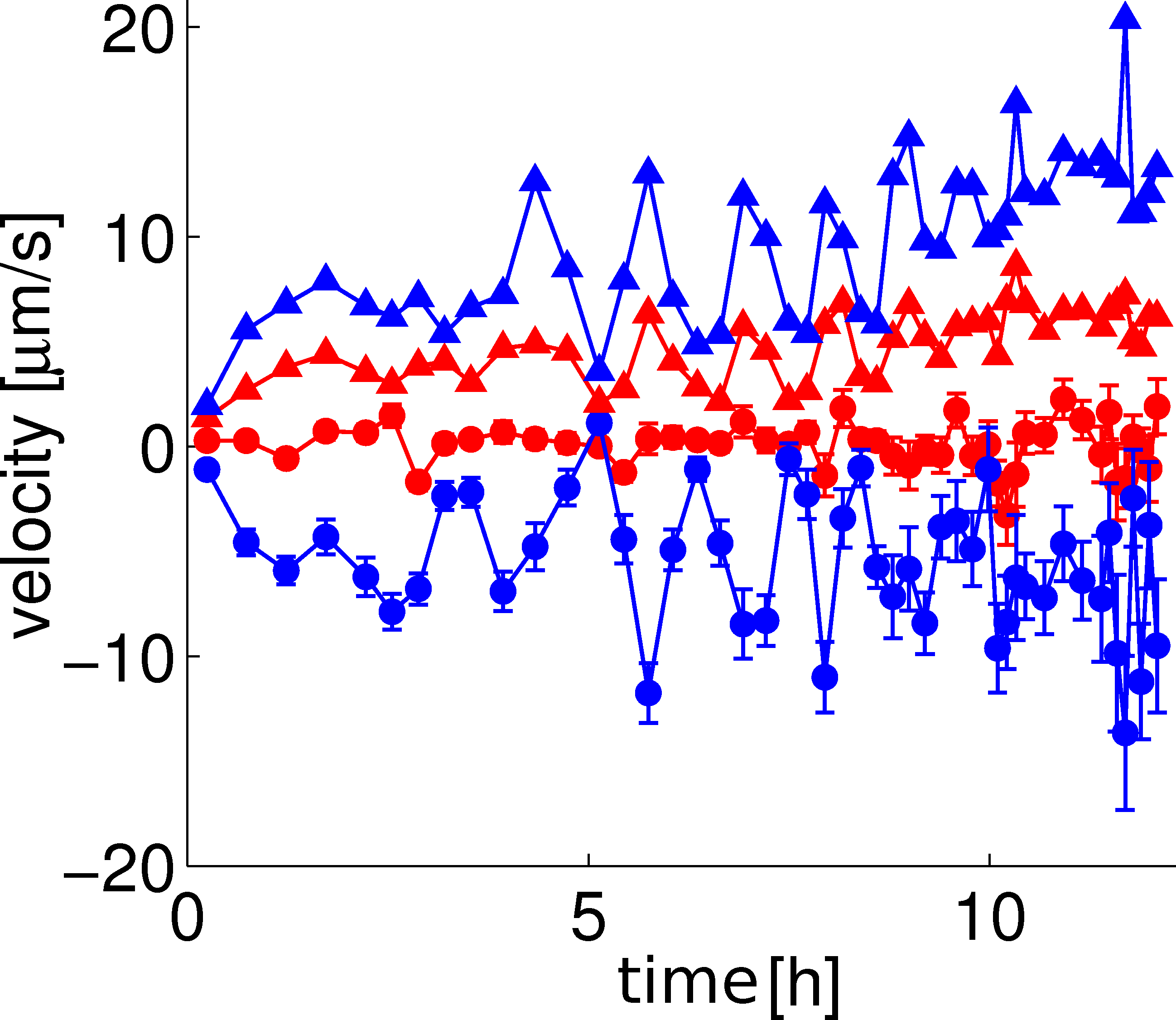}
		\caption{Mean fluid velocity in horizontal (red circles) and vertical (blue circles) direction, 
						and the rms velocity in horizontal (red triangles) and vertical (blue triangles) direction. 
						Data of twelve oscillations are shown, where four consecutive data points correspond to one oscillation. 
						The measurement is performed in the bottom phase with $\xi=1.05\cdot 10^{-5} $s$^{-1}$.}
	\label{fig:oscillatingflow}
\end{figure}

To characterize the time evolution of the flow field (using only non-interpolated grid points), 
the mean vertical and horizontal flow velocity as well as the root mean squared velocity components are calculated. 
For averaging in time, four bins for each oscillation are used (figure \ref{fig:oscillatingflow}). 
The mean horizontal velocity fluctuates around zero as expected for symmetry reasons, 
and the vertical flow velocity is typically negative. 
This can be explained by two effects. 
On the one hand, the upward sedimenting droplet volume has to be balanced by a downward fluid motion to ensure a zero net volume flux. 
On the other hand, there is a large scale convection since the system is heated from outside. 
Close to the walls the compositions of the phases are slightly more separated.
Hence the density of the bulk is higher close to the wall than in the central region of the sample. 
This drives a convection pattern with downward motion close to the wall. 
A coupling of the oscillating droplet size distribution with the flow field produces oscillations in the average vertical flow component and the root mean squared velocities of both components.

\subsection{Oscillation Periods}

\begin{figure}[htb]
	\centering
	\subfigure[\label{fig:periods_top}]
	{\includegraphics[width=\linewidth]{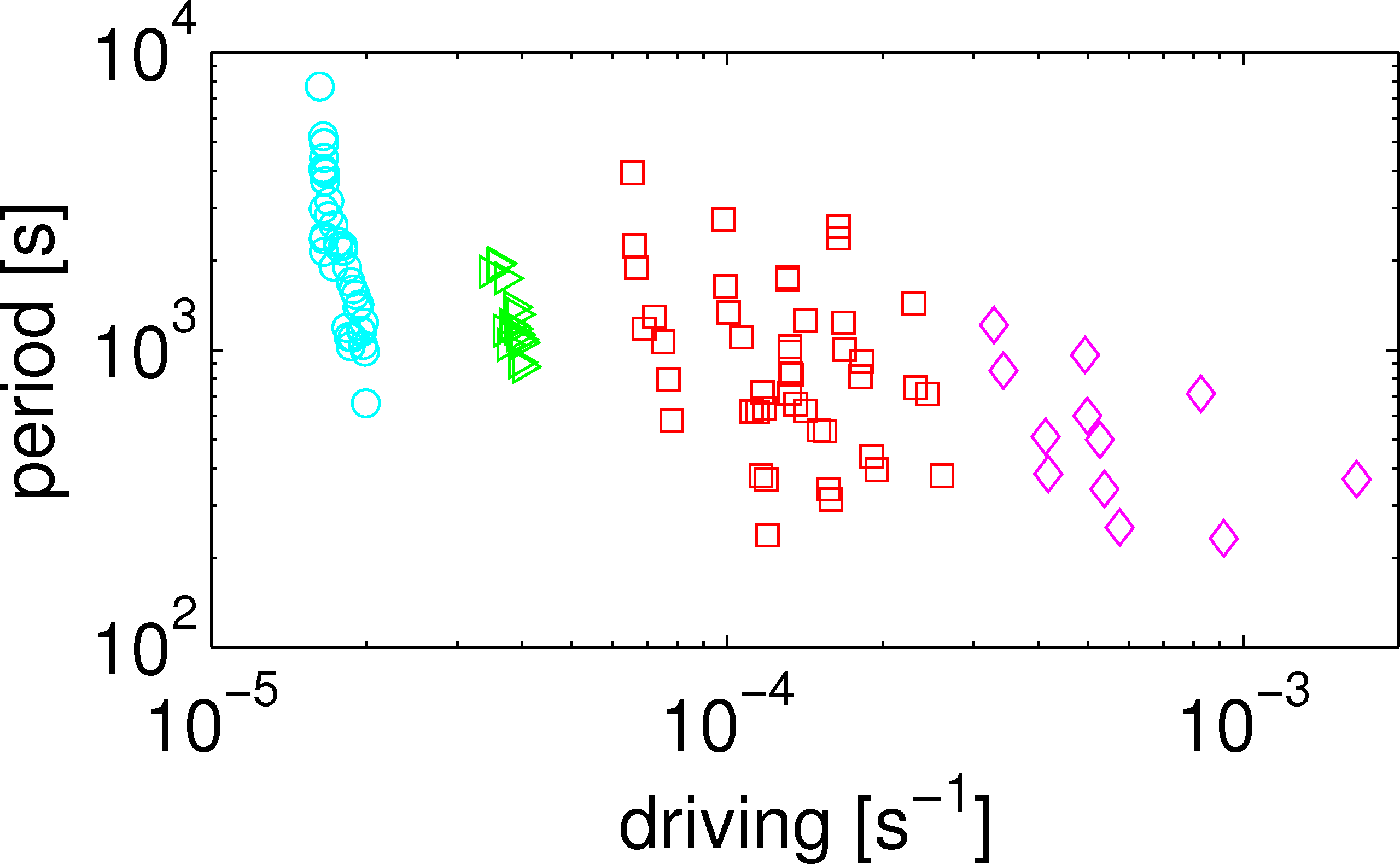}}
	\hfill
	\subfigure[\label{fig:periods_bot}]
	{\includegraphics[width=\linewidth]{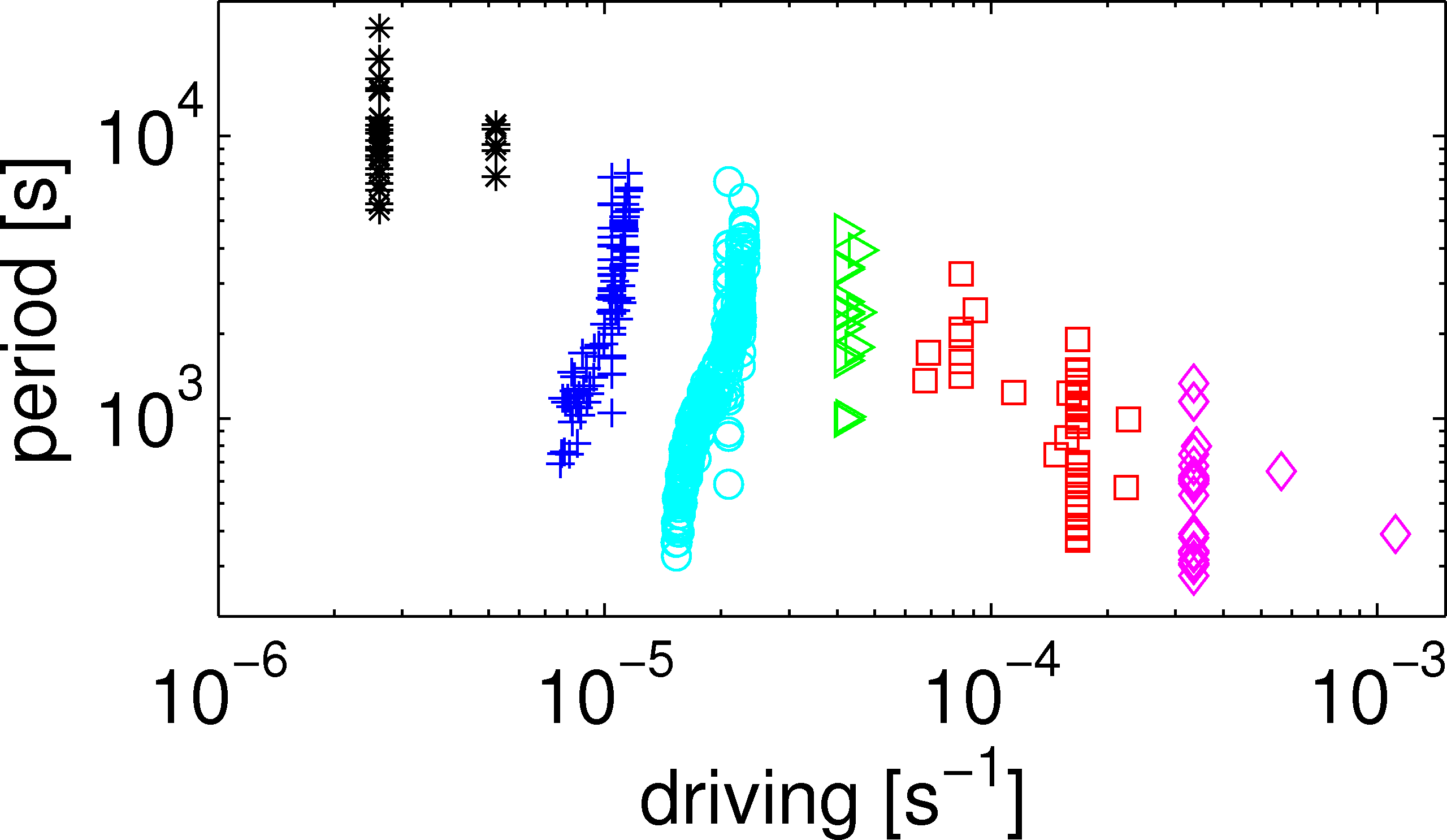}}
	\caption{Oscillation periods $\Delta t$ as a function of the driving rate $\xi$ for the top (a) and the bottom (b) phase.
					 In both phases, the oscillation period is smaller for higher driving rates.
					 In each experimental run, the driving is kept approximately constant. Nevertheless, the oscillation period decreases with increasing temperature.
					 The use of the symbols is described in the text.}
	\label{fig:oscillationperiods}
\end{figure}

In each measurement, the driving rate $\xi$ is kept approximately constant. 
One may therefore expect the system to react with a constant time scale.
However, we find, that during one experimental run, the oscillation period $\Delta t$ decreases with increasing temperature 
(compare figures \ref{fig:sizedistribution} and \ref{fig:oscillatingflow}). 
This trend has been observed in turbidity measurements with other binary mixtures by \cite{Auernhammer2005} and \cite{Vollmer2007} before.
Secondly, the oscillation period depends on the driving rate.
For faster temperature ramps, the oscillation period decreases.
We have measured the oscillation periods of the droplet size distribution for two decades in $\xi$ for the top phase
and almost three decades in $\xi$ for the bottom phase (see figure \ref{fig:oscillationperiods}).
The colors and symbols encode different heating rates $\xi$: 
black stars, $\xi < 6\times 10^{-6}\,{\rm s}^{-1}$; 
blue crosses, $\xi < 1.3\times 10^{-5}\,{\rm s}^{-1}$;
cobalt circles, $\xi < 3\times 10^{-5}\,{\rm s}^{-1}$; 
green triangles, $\xi < 6\times 10^{-5}\,{\rm s}^{-1}$; 
red squares, $\xi < 3\times 10^{-4}\,{\rm s}^{-1}$; 
and magenta diamonds, $\xi > 3\times 10^{-4}\,{\rm s}^{-1}$.
A trend of the period $\Delta t$ with the driving $\xi$ is visible,
but it is masked by the temperature dependence of $\Delta t$.

\section{Discussion \& Conclusions}

To investigate the physics of phase separation we study the water/isobutoxyethanol-system. 
It has very convenient physical properties, a critical point just above room temperature,
an upper miscibility gap, and low vapor pressure. 
Therefore, sample preparation is easy and a broad range of parameters can be studied with a fully automatized setup.
Using Nile Red as a dye and choosing a system with one polar and one organic phase we obtained images with a high fluorescence contrast.

A short arc mercury vapor lamp is suitable for the illumination of the sample.
To follow the evolution of the broad droplet size distribution for several hours, high resolution images (5 Mpixel) are taken at small frame rates (0.3 to 5 Hz).
The resolution of the camera has to be large enough to detect droplet radii, whereas the frame rate must be adapted to the maximal droplet velocities. 
For measuring well resolved flow fields the droplet number density has to be high enough. 
This sets a lower bound to the rate of temperature change, which controls the nucleation of droplets. 
On the other hand, for high droplet number densities, measurements are only possible close to the walls. 
As the droplets are not index matched, clear images can only be obtained at a wall distance up to once or twice the mean distance of droplets. 
Therefore the measurements were conducted 0.3mm away from the wall, i.e. at a distance of about 1.5 times the correlation length calculated from the flow fields.

We presented a particle tracking algorithm based on image processing with MATLAB. 
It detects droplets in a range of 4 to 40 $\mu$m radius (bigger ones can be detected but are very scarce) in a sample field of $1.3\times1.5$mm$^2$. 
The motion of the droplets is decomposed into a sedimentation and an advection term. 
The sedimentation velocity is given by the Stokes velocity of a sphere. 
The advection of droplets is sampled on an Eulerian grid. 
By taking the sedimentation velocity explicitly into account, big droplets can also be used to calculate the flow field, which reduces the noise.
This is possible because the Stokes number is much smaller than one for all droplets in our experiments.
It is not necessary to know in advance the radius at which sedimentation starts to become significant. 
This threshold depends on the specific flow conditions.

There are two advantages by incorporating information on the droplet radius:
First, the radius is a good criterion to identify droplets in subsequent images.
Secondly, the prediction of the droplet position in the next image is improved by knowledge of the radius dependent sedimentation velocity.

The experimental setup and the tracking algorithm are optimized for long measurement times (several hours) and low Reynolds number flow.
Our experimental procedure can be applied to a broad variety of binary mixtures, 
provided an appropriate fluorescent dye for labeling (only) one of the two phases is found and the excitation and emission filters are adapted to it. 
It enables then detailed investigations of the evolution of the system:
The droplet size distribution, Lagrangian particle velocities and Eulerian flow fields can be measured simultaneously for a broad range of heating rates, temperatures and sample geometries. 

To investigate other laminar or turbulent flows a high speed camera has to be used.
Our technique might then be a promising approach to investigate turbulent boundary layers with reactions or phase separation.
Furthermore, one may consider placing an endoscope into the sample, as \cite{Maass2009} have done for example, to measure the size distribution outside the boundary layer. 

A further advantage of the technique is that the simultaneous measurement of particle position and particle size 
allows us to determine the size evolution of individual droplets as they progress across the measurement area. 
This information can be used to obtain information on coalescence rates and collision efficiency of sedimenting droplets:
A preliminary study shows that droplets with radii about $40\mu$m grow on average by $0.5\mu$m/s while they travel across the measurement area. 
This growth amounts to a collision efficiency (see \cite{Pruppacher1997} pp. 569 and \cite{Pinsky1999}) of order unity. 
Further studies are under way to determine how the collision efficiency depends on the radius of the sedimenting droplet. 
This methodology should be of interest to precipitation processes such as rain formation, where the growth of medium sized droplets is believed to be driven by coalescence.

\begin{acknowledgements}
	We are grateful to Wilhelm Hüttner, Konstantin Christou, Kristian Hantke, Alberto de Lozar and Eric Stellamanns for enlightening discussions 
	and experimental tests of several illumination techniques. We thank Doris Vollmer and Günther Auernhammer for advice in designing the setup, 
	choosing the system and developing the experimental protocol. We thank Markus Holzner and Mukund Vasudevan for comments on the manuscript.
	Tobias Lapp acknowledges financial support from Deutsche Forschungsgemeinschaft FOR 1182.
\end{acknowledgements}

\section*{Appendix}

The viscosity of i-BE is measured with an Ubbelohde viscosimeter type 537 10/I made by Schott. The temperature dependence of the viscosity $\eta$ [kg/ms] is fitted by
\begin{equation}
    \eta(T) = A \cdot 10^{\frac{B\cdot (20 - T) - C\cdot (T-20)^2}{T+D}}
    \label{eqn:viscosity_temp}
\end{equation}
 with temperatures expressed in °C. The coefficients are given in table \ref{tab:coeffvisc}, where the values for water are taken from \cite{Weast1988}.
\begin{table}[h]
      \centering
      \begin{tabular}{lllll}
				  & $A [kg/ms]$	& $B$	& $C$	& $D$		\\
	  \hline
	  water & $1.002\cdot 10^{-3}$ & 1.3272 & 0.001053 & 105\\
	  i-BE  & $3.36\cdot 10^{-3}$  & 1.730  & 0.001    & 108
      \end{tabular}
      	  \caption{Fit coefficients for the viscosity of water (\cite{Weast1988}) and i-BE, defined by equation (\ref{eqn:viscosity_temp}).}
	  \label{tab:coeffvisc}
\end{table}

To interpolate the viscosities for a mixed phase of given mass fraction $\Phi$ we use the composition-dependent viscosities at 25°C 
for a homogeneous mixture in the single phase regime given in \cite{Menzel2003}. The data is fitted with a fifth order polynomial 
\begin{equation}
	\begin{aligned}
		\eta(\Phi,{\rm T}=25^{\circ}{\rm C}) = -40.66\Phi^5 +103.44\Phi^4 -100.32\Phi^3 \\
										 +39.35\Phi^2 +0.17\Phi +0.91.
	\label{eqn:viscfit}
	\end{aligned}
\end{equation}

\begin{figure}[tb]
	\centering
	\includegraphics[width=0.80\linewidth]{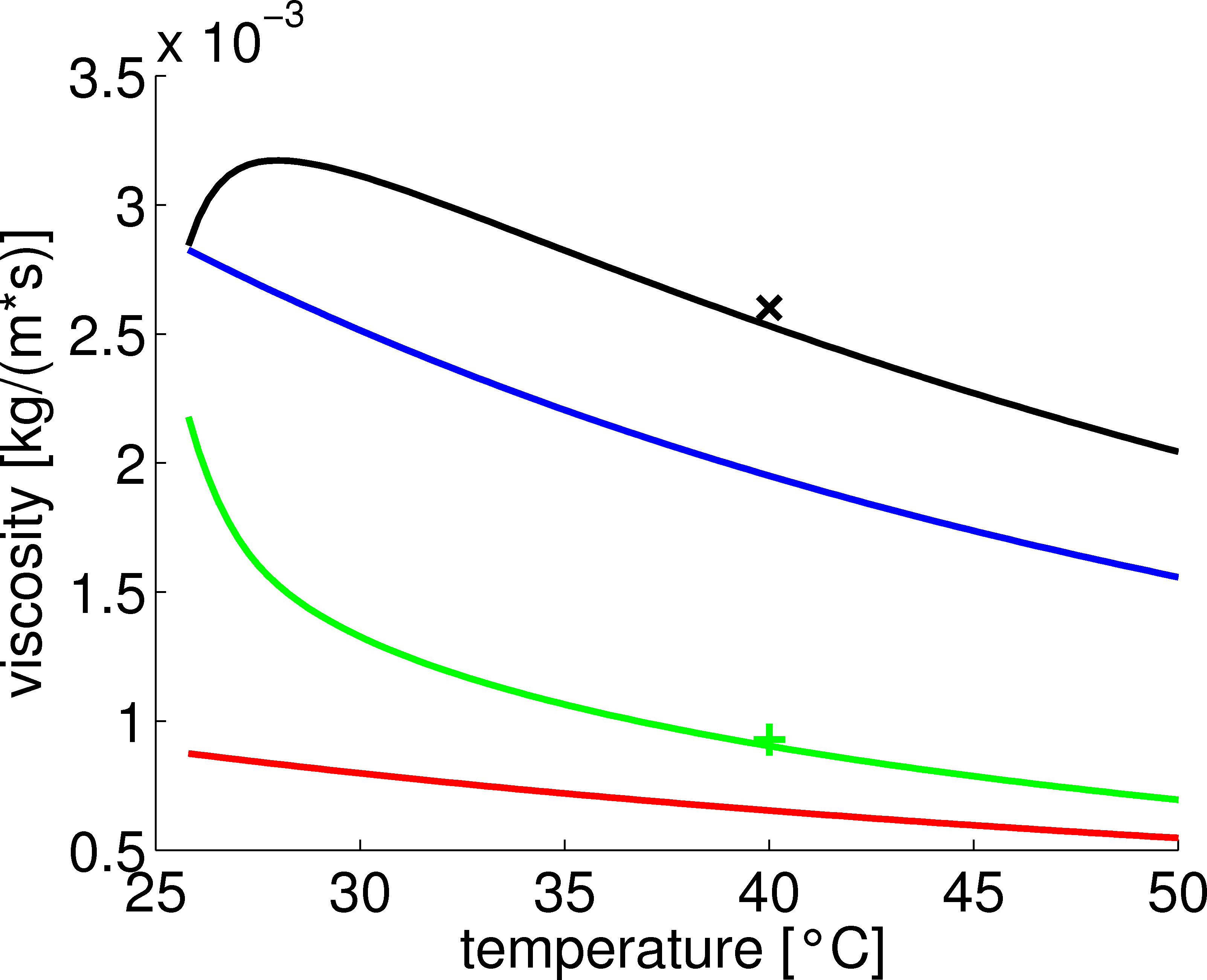}
	\caption{From top to bottom: Viscosity of the top phase, isobutoxyethanol, the bottom phase and water, using equations (\ref{eqn:viscosity_temp},\ref{eqn:etafinal}). 
					 At 40°C the viscosity of the two mixed phases was measured with an Ubbelohde viscosimeter.}
	\label{fig:viscosity}
\end{figure}

Assuming that the coefficients of interpolation are not changing substantially in the temperature range of our measurements, a rescaled viscosity $\tilde{\eta}(\Phi)$ is defined. 
It only depends on the composition $\Phi$
\begin{equation}
\eta(\Phi,T) = \tilde{\eta}(\Phi)\cdot \eta_{iBE}(T) + (1-\tilde{\eta}(\Phi))\cdot \eta_{H_2O}(T).
\label{eqn:etafinal}
\end{equation}
The viscosities of the two phases are shown in figure \ref{fig:viscosity} as a function of temperature.
To check the strong assumption entering this interpolation, we measured the viscosity of the two phases at $T=40$°C.
For both phases it followed the prediction of equation (\ref{eqn:etafinal}) to within 2\%.
This is sufficiently accurate for our means.

\end{document}